\documentclass[aps,prx,twocolumn,showpacs,superscriptaddress,longbibliography]{revtex4-2}
\usepackage{amsmath, amsfonts, amssymb,graphicx}
\usepackage{graphicx,epstopdf}
\usepackage{gensymb}
\usepackage[dvipsnames]{xcolor}
\usepackage{booktabs}
\usepackage{adjustbox}
\usepackage{multirow}
\usepackage{geometry}
\newcommand{\be}{\begin{equation}}
\newcommand{\ee}{\end{equation}}
\newcommand{\bea}{\begin{eqnarray}}
\newcommand{\eea}{\end{eqnarray}}
\newcommand{\bse}{\begin{subequations}}
\newcommand{\ese}{\end{subequations}}

\usepackage{multibib}
\usepackage{color}
\usepackage[colorlinks,bookmarks=false,citecolor=darkblue,linkcolor=red,urlcolor=blue]{hyperref}

\definecolor{darkred}{rgb}{0.7,0.0,0.0}

\definecolor{darkblue}{rgb}{0,0.02,0.45}

\definecolor{darkgreen}{rgb}{0.02,0.45,0.0}

\definecolor{violet}{rgb}{0.8,0.2,0.6}

\begin{document}

\preprint{APS/123-QED}

\title{Evidence for magnetoelastic coupling and chiral magnetic ground state in quasi-van der Waals \textit{tr}-Cr$_{1.22}$Te$_{2}$}


\author{S. M. Hossain}
\affiliation{Department of Physics, Shiv Nadar Institution of Eminence, Gautam Buddha Nagar, Uttar Pradesh 201314, India}
 
\author{B. Rai}
\address{S. N. Bose National Centre for Basic Sciences, Salt Lake City, Kolkata 700106, India}

\author{P. R. Baral}
\address{PSI Center for Neutron and Muon Sciences, 5232 Villigen PSI, Switzerland}

\author{O. Zaharko}
\address{PSI Center for Neutron and Muon Sciences, 5232 Villigen PSI, Switzerland}

\author{N. Kumar}
\address{S. N. Bose National Centre for Basic Sciences, Salt Lake City, Kolkata 700106, India}

\author{A. K. Bera}
\email{akbera@barc.gov.in}
\address{Solid State Physics Divison, Bhabha Atomic Research Center, Mumbai 400085, India}
\address{Homi Bhabha National Institute, Anushaktinagar, Mumbai 400094, India}

\author{M. Majumder}
\email{mayukh.majumder@snu.edu.in}
\affiliation{Department of Physics, Shiv Nadar Institution of Eminence, Gautam Buddha Nagar, Uttar Pradesh 201314, India}

\date{\today}           

\begin{abstract}
Trigonal (\textit{tr})-Cr$_{1+\delta}$Te$_{2}$ is a well-known ferromagnetic material that has recently drawn much attention due to the discovery of zero-field skyrmion state, unusual anomalous Hall effect, topological Hall effect, and topological Nernst effect. This quasi-van der Waals (vdW) layered material with intercalated Cr atoms possesses many peculiar features that depend on the amount of Cr intercalation, although the microscopic magnetic ground state is still elusive. We reveal the structural and magnetic properties of \textit{tr}-Cr$_{1.22}$Te$_{2}$ by low-temperature x-ray diffraction, magnetization, temperature-dependent Raman spectroscopy, and single crystal neutron diffraction studies. Magnetization measurements under small applied magnetic field indicate two successive magnetic transitions, one from a ferromagnetic (FM) state to an antiferromagnetic (AFM) state (\textit{T}$_\mathrm{C}=197$~ K), and second from AFM to a paramagnetic state (\textit{T}$_\mathrm{N}=211$~ K). The FM transition is sharp with a strong presence of magnetoelastic coupling, but is not accompanied by any structural phase transition. The magnetic structure obtained from zero-field single crystal neutron diffraction reveals that the Cr1 and Cr2 moments are ferromagnetically aligned along the \textit{c}-axis, while the Cr3 and intercalated Cr4 atoms induce an AFM component in the \textit{ab}-plane leading to an umbrella-like spin structure which possesses a finite spin chirality. The presence of a finite spin chirality is responsible for the observation of the topological Hall effect (THE).
\end{abstract}                       

\maketitle

\section{Introduction} 
The general belief was that long-range magnetic order is not guaranteed in 2D Heisenberg materials because of the enhancement of low-energy spin waves, which destroys magnetic order at finite temperature as proposed by the Mermin-Wagner theorem~\cite{PhysRevLett.17.1133}. However, the recent discovery of long-range magnetic ordering in monolayers of two-dimensional (2D) vdW materials has opened new directions in condensed matter physics for both application and fundamental interests \cite{RevModPhys.95.035002, 10.1093/acprof:oso/9780199581931.001.0001}. The presence of anisotropic interactions (Ising or XY type, etc.) is one of the possible reasons for the development of long-range magnetic order, which opens a gap in spin wave excitations~\cite{Kim2024, PhysRevX.11.031047, Jana2024}. Among all the vdW systems, transition metal dichalcogenides are of particular interest due to a large pool of compounds to choose from, structural flexibility, and the ability to intercalate atoms into the vdW gaps~\cite{doi:10.1021/acs.chemrev.7b00618}. Very recently, a self-intercalated trigonal compound with composition Cr$_{1+\delta}$Te$_{2}$ where $\delta\approx$ 0.33 showed interesting magnetic properties. Lorentz transmission electron microscopy uncovered Néel-type skyrmionic phase in the 100-200 K temperature range, even in zero magnetic fields \cite{Saha2022, Rai2025}. The lack of inversion symmetry of the crystal structure promotes the Dzyaloshinskii-Moriya interaction (DMI) in addition to the Heisenberg exchange, which stabilises the skyrmion state~\cite{doi:10.1126/science.1166767, PhysRevB.81.041203, doi:10.1126/science.1214143,PhysRev.120.91, Dzyaloshinsky1958ATT}. Many peculiar magnetic ground states can be stabilised upon changing the amount of Cr intercalation, resulting in a wide range of magnetic and transport phenomena~\cite{PhysRevB.110.L060405, GUPTA2025101627}. Recently, a finite topological Hall conductivity has been measured in \textit{tr}-Cr$_{1+\delta}\mathrm{Te}_{2}$ with (0.2 $\le \delta \le$ 0.33)~\cite{Rai2025, 10.1063/1.5143387, chowdhury2024suppressionintrinsichalleffect}. A non-coplanar spin configuration with spin chirality was proposed to promote a finite topological Hall conductivity (THC) \cite{PhysRevLett.102.186602, Nagaosa2013}. So far, little initiative has been taken to understand the magnetic structure of this family of compounds with different Cr and Te concentrations using neutrons as a microscopic tool~\cite{Andresen1963AND, Huang2008AND}. Moreover, the knowledge of the magnetic ground state could shed light on the factors generating the THE in this system.

\begin{figure*}
      \centering
      \includegraphics[width=1\textwidth]{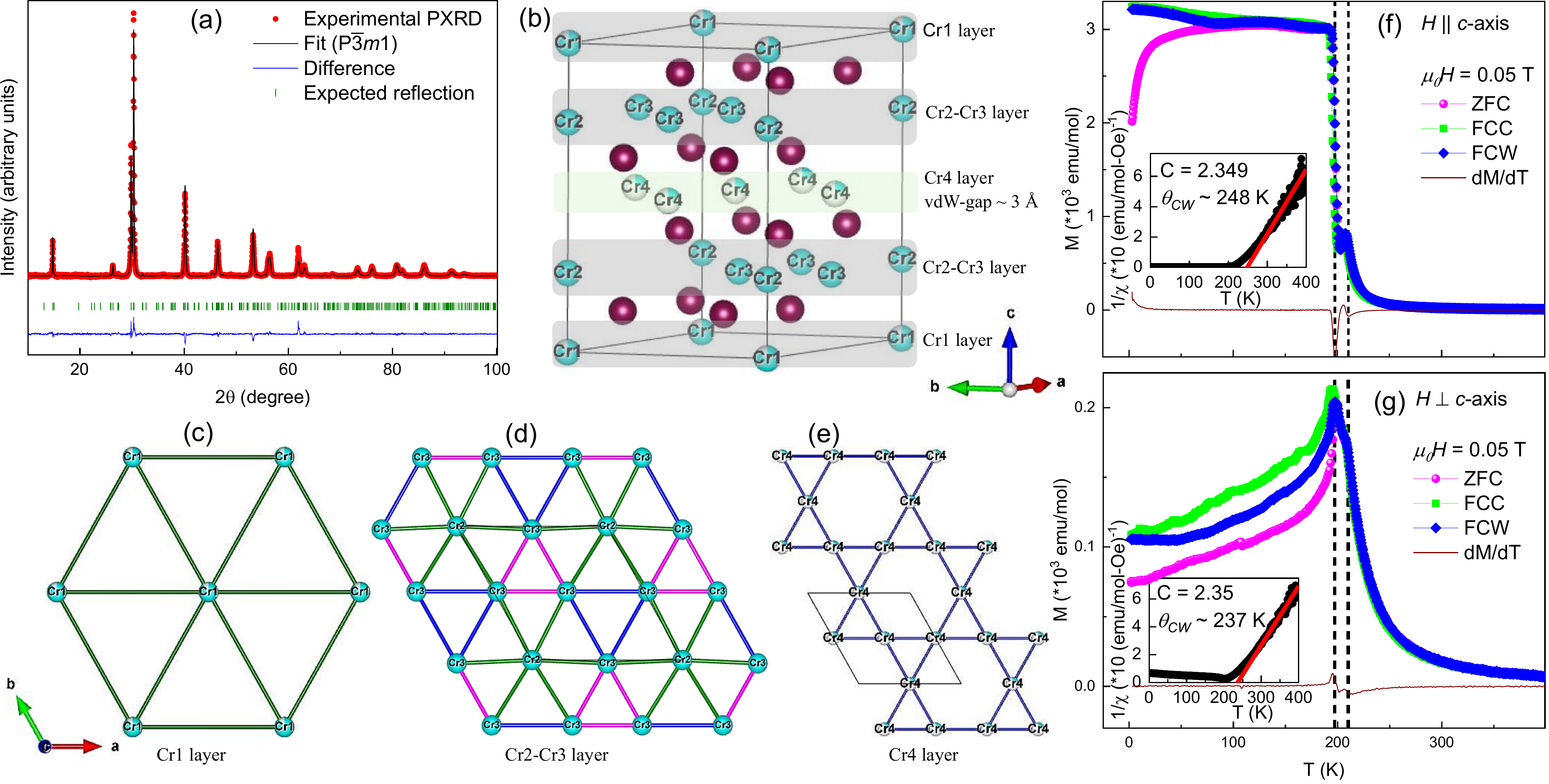}
      \caption{(a) The Rietveld refinement of room-temperature PXRD pattern of \textit{tr}-Cr$_{1.22}$Te$_{2}$. (b) The crystal structure of \textit{tr}-Cr$_{1.22}$Te$_{2}$ with cyan and bordeaux colour spheres indicating the Cr and Te atoms, respectively. The distance between Cr1-Cr2 layers is 3.067~\AA, and the distance between adjacent Cr2-Cr3 layers is: 5.84~\AA between Cr2-Cr2 and 6.28~\AA for Cr3-Cr3. (c-e) The different Cr sites make geometrically frustrated lattices (triangular and kagome) arrangements. (f) and (g) represent magnetization measurement of \textit{tr}-Cr$_{1.22}$Te$_{2}$ with the magnetic field applied parallel and perpendicular to the crystallographic \textit{c}-direction, respectively. The solid bordeaux colour line represents the first derivative (d$M$/d$T$) of the FCW curve with the \textit{T}$_\mathrm{C}$ and \textit{T}$_\mathrm{N}$ at 197~K and 211~K, respectively. Inset figures show the Curie-Weiss fit.}
      \label{fig: XRD_magnetization}
\end{figure*} 

In this work, we investigated in detail low-temperature crystallographic and magnetic properties of quasi-vdW \textit{tr}-Cr$_{1+\delta}$Te$_{2}$ ($\delta\approx$ 0.22) intermetallic compound, which hosts Néel-type Skyrmion phase even at zero applied magnetic field \cite{Rai2025}. From detailed magnetization measurements, we identified two successive magnetic orderings along the \textit{c}-axis, whereas only AFM ordering perpendicular to the \textit{c}-direction is pronounced. Temperature-dependent powder x-ray diffraction (PXRD) and Raman spectroscopic studies indicated the presence of strong spin-phonon coupling (SPC). Furthermore, the ground state magnetic structure was determined from single crystal neutron diffraction. Spins located at the edges of the unit cell are completely aligned along the \textit{c}-axis, whereas the spins located inside the unit cell have chiral components leading to an umbrella-like spin configuration. Thus, the ground state magnetic structure of \textit{tr}-Cr$_{1.22}$Te$_{2}$ obtained from neutron diffraction is a canted ferromagnetic spin arrangement with spin chirality which is responsible for the observation of THE in \textit{tr}-Cr$_{1+\delta}$Te$_{2}$ ($\delta \approx$ 0.22).

\begin{figure}
      \centering
      \includegraphics[width=0.47\textwidth]{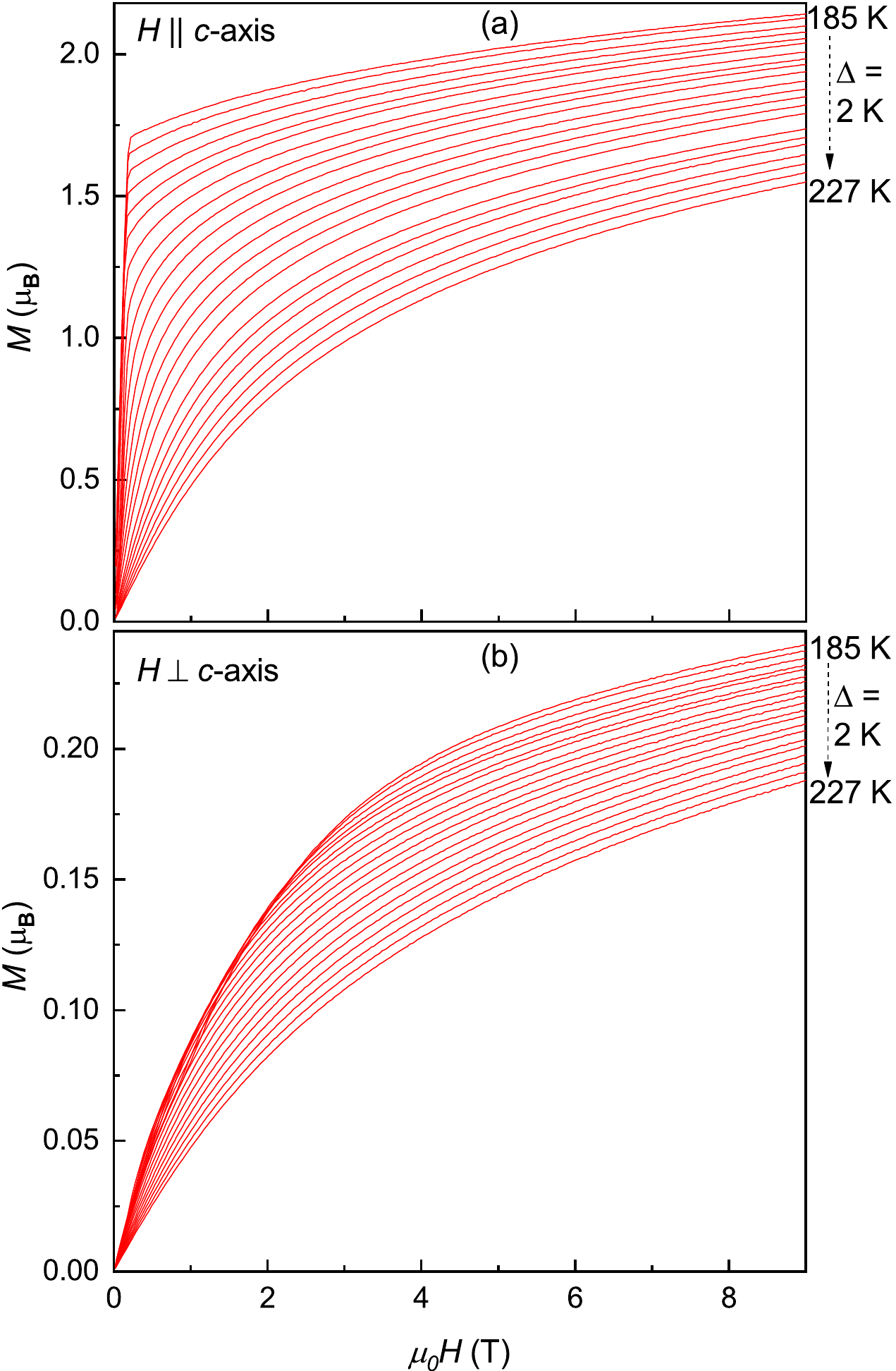}
      \caption{(a) and (b) are isothermal magnetization curves (magnetisation scan at a constant temperature by sweeping the applied magnetic field) $M$ vs $H\parallel$ \textit{c} and $H\perp$ \textit{c}, respectively in the temperature range of 185~K to 227~K.}
      \label{fig:  Magnetization_2}
\end{figure}

\section{Experimental Methodology} 
For this study, high-quality plate-like millimetre-sized single crystals of \textit{tr}-Cr$_{1+\delta}$Te$_{2}$ ($\delta \approx$ 0.22) were grown by the self-flux method (reported elsewhere \cite{Rai2025}). Powder x-ray diffraction (PXRD) patterns were collected on the grounded crystals using a Rigaku SmartLab diffractometer with x-ray Cu K$_\alpha$ source within the 2$\theta$ range 10$\degree$-100$\degree$ with a temperature step of 10~K from room temperature (RT) down to 20~K. The temperature-dependent Raman spectroscopy experiment was carried out on an as-grown bulk single crystal sample of \textit{tr}-Cr$_{1.22}$Te$_{2}$ using a LabRam HR Evolution Raman spectrometer (HORIBA France SAS-532 nm laser) in the temperature range of 300~K to 88~K.

The magnetization measurements were performed parallel (crystallographic \textit{ab}-plane "flat surface" parpendicular to the applied magnetic field) and perpendicular (crystallographic \textit{ab}-plane "flat surface" parallel to the applied magnetic field) to the crystallographic \textit{c}-axis using a commercially available physical property measurement system (PPMS) by Quantum Design, which measures in the temperature range from 400~K to 2~K. The magnetization vs temperature dependence (\textit{M} vs \textit{T}) was measured with an applied field of 0.05 T in zero-field cooling (ZFC), field-cooled cooling (FCC), and field-cooled warming protocol (FCW). The isothermal magnetization as a function of applied field (\textit{M} vs \textit{H}$\parallel$\textit{c}) and \textit{H}$\perp$\textit{c} curves were obtained at a constant temperature by allowing the sample to reach thermal equilibrium before starting the measurement in the magnetic field range of 0 to 9 T.

Zero-field single crystal neutron diffraction measurements were performed over a 7-300~K temperature range using the ZEBRA single crystal neutron diffractometer at Swiss Spelletion Neutron Source (SINQ) at Paul Scherrer Institute (PSI), Switzerland. A crystal of size 1×5×9 mm$^3$ with a weight of approximately 100~mg [shown in the inset of Fig. 6(a)] was mounted along the \textit{c}- direction vertically into a custom-made Al-sample holder.  For the low-temperature measurements, the sample with the Al-holder was mounted in a closed-cycle refrigerator attached to the 4-circle Eulerian cradle. A neutron beam of the wavelength 1.383~\AA was selected by the Ge-monochromator in the bisecting geometry. The single crystal neutron diffraction datasets were analysed using the FULLPROF SUITE software package~\cite{RODRIGUEZCARVAJAL199355}.

\section{Results and discussion}

\subsection{Crystal structural properties} 
Fig. \ref{fig: XRD_magnetization}(a) is the Rietveld refinement of the RT PXRD pattern of quasi-vdW compound \textit{tr}-Cr$_{1.22}$Te$_{2}$ under P$\bar{3}$m1 space group (details have been discussed in ~\cite{Rai2025}). 
The obtained crystal structure (Fig.  \ref{fig: XRD_magnetization}(b)) shows a layered structure consisting of  Cr and Te atoms (marked as cyan and bordeaux colour solid spheres, respectively). There are a total of four such Cr-Te layers per unit cell along the \textit{c}-axis for Cr$_{1.22}$Te$_2$. Out of four layers, the top two Cr-Te layers are inverted with respect to the bottom two Cr-Te layers, resulting in a vdW gap ($\sim$3~\AA). The top and bottom-most layers are composed of nearly fully occupied (89~$\%$ occupied) Cr1 atoms. The inner two layers are composed of fully occupied Cr2 and Cr3 sites. Our analyses reveal that the Cr4 atoms get intercalated within the vdW gap with a partial occupancy (29~$\%$) and are shown by white-cyan spheres.

\begin{figure*}
      \centering
      \includegraphics[width=1\textwidth]{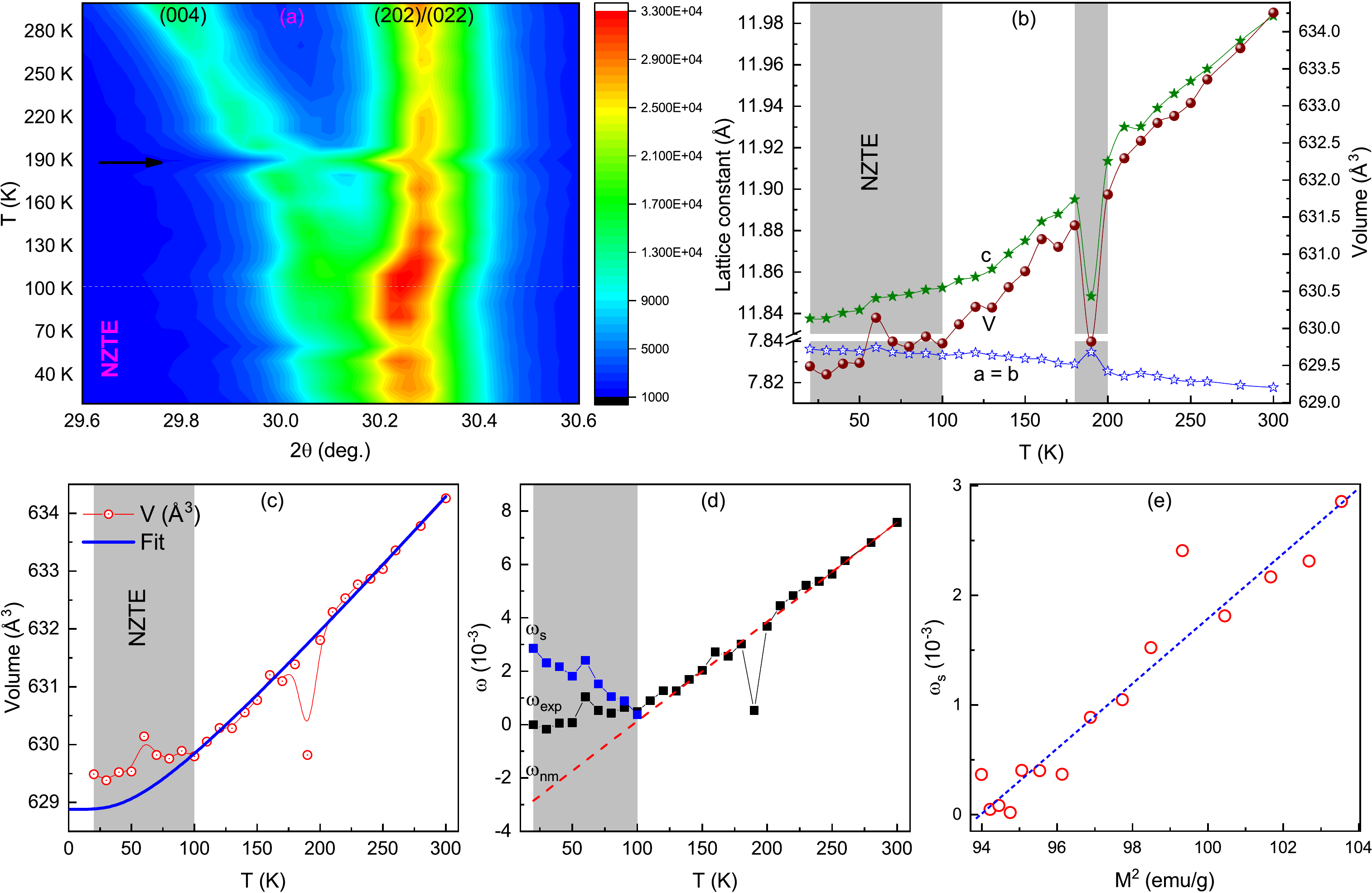}
      \caption{(a) Colour map of the maximum intensity Bragg peak observed in PXRD patterns with temperature evaluation. (b) The temperature evaluation of lattice parameters obtained after refinement shows opposite thermal expansion behaviour for the in- and out-of-the basal plane direction of \textit{tr}-Cr$_{1.22}$Te$_{2}$. (c) The phonon contribution (blue line) associated with the change in the unit cell volume (open red circles) in the high-temperature region follows nicely the Debye-Gr\"{u}nesien approximation (Eq. \ref{Phonon}), whereas in the low-temperature region, the presence of nearly zero thermal expansion (NZTE) causes deviation from it. (d) Black and blue solid squares represent observed and the anomalous thermal expansion (shown in the dashed red line obtained by Eq. \ref{TE}) of \textit{tr}-Cr$_{1.22}$Te$_{2}$. (e) The anomalous thermal expansion behaviour follows a linear relation, Eq. \ref{ATE}, with the measured moment at corresponding temperatures.}
      \label{fig: Low-temperature XRD}
\end{figure*}

\subsection{Magnetic properties}
Fig. \ref{fig: XRD_magnetization}(f) and Fig. \ref{fig: XRD_magnetization}(g) show the temperature dependence of the magnetization curves for a field 0.05~T applied parallel and perpendicular to the crystallographic \textit{c}-axis, respectively.

The magnetization along the \textit{c}-axis indicates an AFM ordering at 211~K (\textit{T}$_\mathrm{N}$) estimated from the first derivative (d\textit{M}/d\textit{T}), as shown by the solid maroon line in Fig. \ref{fig: XRD_magnetization}(f). With reducing the temperature, a sharp enhancement of magnetization is observed, followed by a saturating behaviour at further low temperatures, indicating a ferromagnetic ordering with \textit{T}$_\mathrm{C}\approx$ 197~K (estimated by following the same procedure as \textit{T}$_\mathrm{N}$). Note that a bifurcation below 100 K between ZFC and FC curves indicates the ferromagnetic domain wall pinning effect\cite{Dara_2023, MONDAL201927, PhysRevB.100.024434}. Moreover, the dominance of FM state below \textit{T}$_\mathrm{C}$ is also supported by the finite and large spontaneous magnetization observed in the \textit{M} vs \textit{H} isotherms below  \textit{T}$_\mathrm{C}$ (Fig. \ref{fig: Magnetization_2}(a)). \textit{T}$_\mathrm{C}$ and \textit{T}$_\mathrm{N}$ merge with increasing fields, which is expected because the magnetic field usually tends to increase the FM ordering temperature and reduce the AFM ordering temperature. Thus, the measurement at low field (0.05~T) was important to observe a clear AFM ordering at \textit{T}$_\mathrm{N}$, which is often not visible in other studies~\cite{Saha2022, PhysRevMaterials.7.044406, PhysRevB.100.245114, MONDAL201927}. The observation of \textit{T}$_\mathrm{N}$ could be due to the ordering of the intercalated Cr4 atoms, as the magnetization value associated with \textit{T}$_\mathrm{N}$ is much smaller compared to \textit{T}$_\mathrm{C}$. Note that in the neutron diffraction study (discussed later), no anomaly is observed at \textit{T}$_\mathrm{N}$, which further supports that a small moment fraction is associated with the ordering at \textit{T}$_\mathrm{N}$. However, further detailed investigations are required to determine the origin of the AFM ordering.

Fig. \ref{fig: XRD_magnetization}(g) presents \textit{M} vs \textit{T} data for \textit{H} perpendicular to the crystallographic \textit{c}-direction showing AFM behaviour with no trend of saturation in the isothermal \textit{M} vs \textit{H} curves up to 9~T (Fig. \ref{fig: Magnetization_2}(b)). Moreover, a large difference in the value of magnetization from \textit{M} vs \textit{T} plots for field applied parallel and perpendicular to the \textit{c}-axis (Fig. \ref{fig: XRD_magnetization}(f) and Fig. \ref{fig: XRD_magnetization}(g)) shows a significant magnetocrystalline anisotropy, which may arise from the presence of spin-orbit coupling (SOC) associated with Te~\cite{Rai2025}. The SOC may also induce anisotropy in the \textit{ab}-plane. Thus, to study the in-plane anisotropy, an angular dependence of magnetization measurement within the \textit{ab}-plane is needed, which is out of the scope of the present work.

\begin{figure*}
      \centering
      \includegraphics[width=0.95\textwidth]{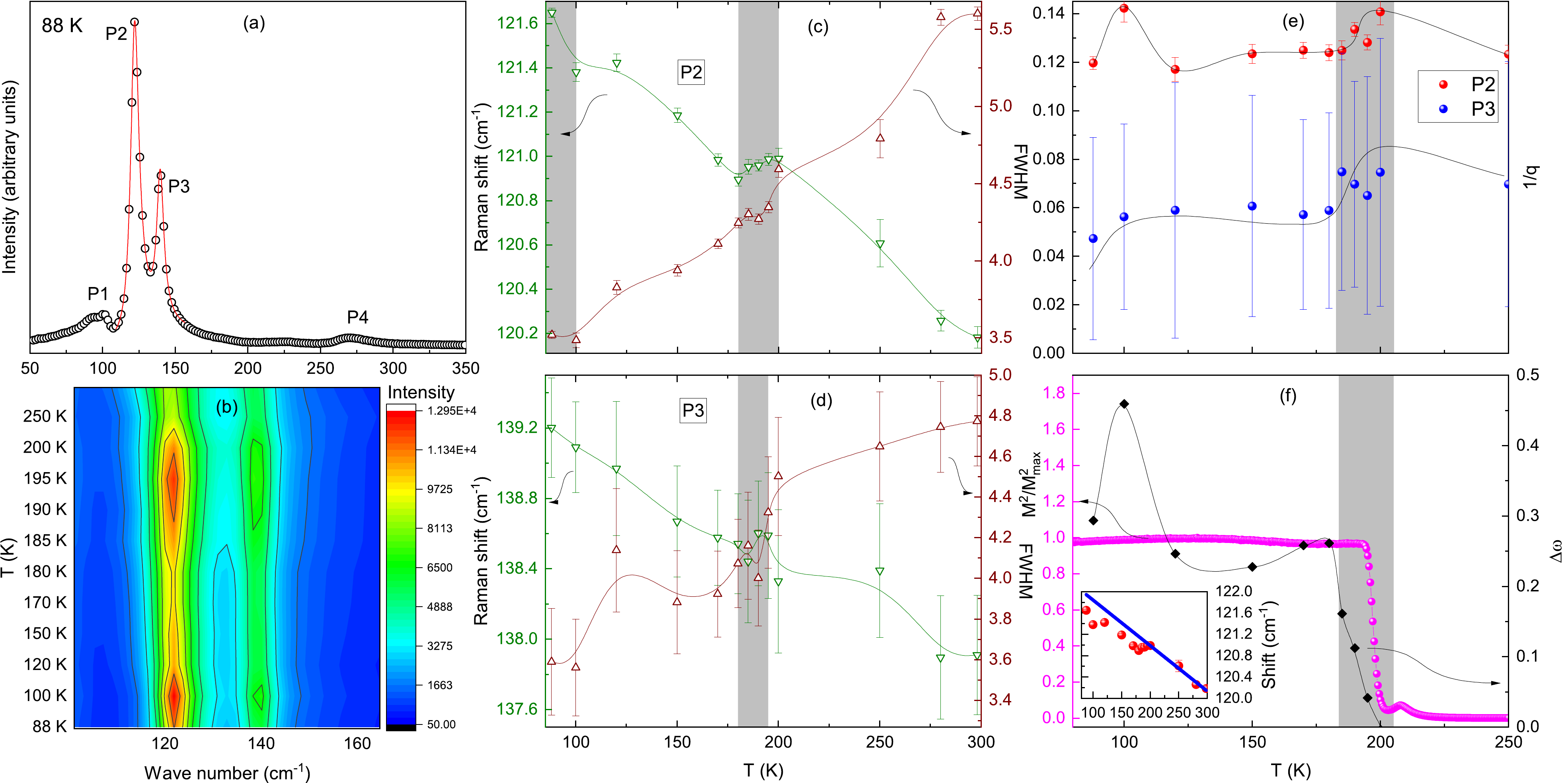}
      \caption{(a) Raman intensity as a function of wave number measured at 88 K. Four major peaks at 100~cm$^{-1}$, 121~cm$^{-1}$, 138~cm$^{-1}$, and 270~cm$^{-1}$ are observed, denoted as P1, P2, P3, and P4, respectively. (b) The temperature-dependent contour plot of the maximum intensity peaks P2 and P3 with respect to wave number. (c)-(d) are the Raman shift and FWHM obtained by fitting Eq. \ref{Fano} of P2 and P3, respectively, as a function of temperature. (e) Fano asymmetry of both the peaks, red and blue filled spheres corresponding to P2 and P3, respectively. The black line is a guide to the eye. (f) Normalized local magnetic moments as a function of temperature (magenta-filled spheres) and the $\Delta\omega$ as a function of temperature show similar behaviour. The inset red-filled circles are the Raman shift of P2, and the solid blue line is a fit to the Eq. \ref{Anharmonic}.}
      \label{fig: LTRaman}
\end{figure*} 

The susceptibility ($\chi=M/\mu_0H$) data above 300~K has been fitted with Curie-Weiss law
\begin{equation}
\chi=\frac{C}{T-\theta_{CW}}
\end{equation}
shown as a solid red line in the inset of Fig. \ref{fig: XRD_magnetization}(c) and (g). The parameters obtained from the fitting for the magnetic field parallel to the \textit{c}-axis are Curie-Weiss temperature $\theta_{CW}\approx$ 250~K and the Curie constant $C=2.349$ cm$^3$Kmol$^{-1}$. These parameters for magnetic field perpendicular to \textit{c}-axis are $\theta_{CW}\approx$ 240~K,  $C=2.35$ cm$^3$Kmol$^{-1}$. The positive sign of $\theta_{CW}$ indicates that the nature of exchanges is predominantly ferromagnetic. The calculated effective moment using $\mu_{eff}=\sqrt{8C}$ from the Curie-Weiss fits in the paramagnetic state is 3.9 $\mu_B$/Cr, which closely resembles the theoretical value of 3.87 $\mu_B$ expected for Cr$^{3+}$. However, the ordered moment of Cr atoms obtained from saturation magnetization parallel to the \textit{c}-axis from \textit{M} vs \textit{H} measurement is about 2.5 $\mu_B$ at 50 K, consistent with other reports \cite{PhysRevB.106.094407}.  This is much smaller than the expected value for spin, $S={3/2}$ of Cr$^{3+}$.

Note that the magnetic ordering temperature \textit{T}$_\mathrm{C}$ or even \textit{T}$_\mathrm{N}$ is lower than the $\theta_\mathrm{CW}$. This indicates a possible presence of frustration in this compound. The appearance of frustration is expected as all the Cr sites (Cr1, Cr2, Cr3, and Cr4) are arranged in a geometrically frustrated edge-shared triangular and corner-shared triangular (kagome lattice) structure, respectively, as shown in Fig. \ref{fig: XRD_magnetization}(c-e). Moreover, the deviation of 1/$\chi$ from linearity below 300~K indicates the development of short-range correlation~\cite{PhysRevB.94.014426}.

To check the degree of itineracy, we have evaluated the \textit{Rhodes Wohlfarth} Ratio (RWR), which is a ratio between the moment in the paramagnetic state ($\mu_{c}$) to the spontaneous magnetization ($\mu_{s}$). Here, $\mu_c$ can be obtained from $\mu_{eff}^2=\mu_c(\mu_c+2)$, where $\mu_{eff}$ is the effective paramagnetic moment, and $\mu_s$ is the spontaneous magnetisation in the ground state \cite{doi:10.1143/JPSJ.55.3553}. The RWR for \textit{tr}-Cr$_{1.22}$Te$_{2}$ obtained as $\mu_{c}/\mu_{s} >1$ from the Stoner model indicates the presence of itinerant spins here as expected. This is further attributed to the itinerant nature and significant spin fluctuations, promoting suppression of the saturation moment as discussed above for the \textit{M} vs \textit{H} measurements.
 
\begin{figure}
      \centering
      \includegraphics[width=0.47\textwidth]{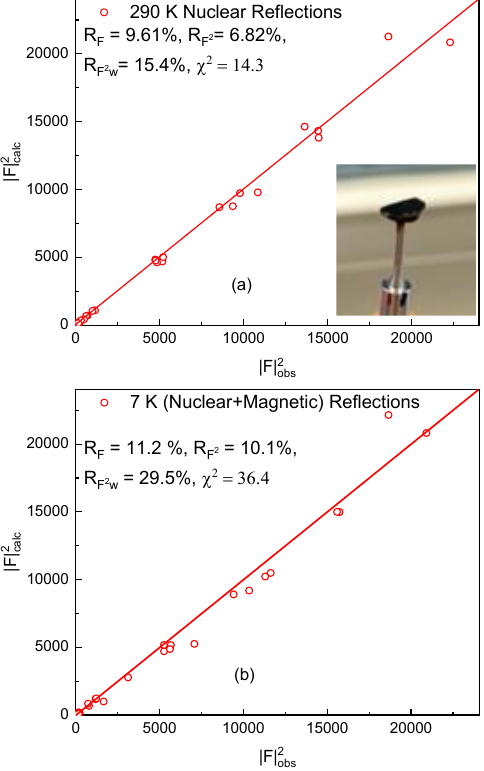}
      \caption{(a) and (b) show the observed vs calculated intensity (square of the structure factor) of the nuclear (290~K) and (nuclear+magnetic) at 7~K Bragg reflections, respectively. The agreement factors are also indicated therein.}
      \label{fig: Neutron refinement}
\end{figure}

\subsection{Low-temperature structural properties and magnetoelastic coupling} 
\subsubsection{Temperature-dependent PXRD study}
A temperature-dependent powder x-ray diffraction (PXRD) study was performed on powdered single crystals. The data were collected with a temperature interval of 10 K, starting from 300 K to 20 K. The Rietveld refinement of all diffraction patterns confirm the trigonal symmetry of the crystal structure with temperature evolution. All observed Bragg peaks were fitted well with the P$\bar{3}$m1 space group (No. 164). No additional Bragg peaks emerge in the studied temperature range, particularly in high diffraction angles, confirming the absence of a structural transition. However, a modification in the positions of diffraction peaks has been observed by closer inspection of the diffraction patterns at base temperature (20~K) and RT (300~K). The major change can be observed in the two strong peaks  (004) and (202)/(022). The temperature evaluation of these two peaks plotted as a function of temperature (from 300~K to 20~K) in the 2$\theta$ range of 29.5$\degree$ to 30.5$\degree$ shows a Y-shaped contour pattern in Fig. \ref{fig: Low-temperature XRD}(a). This implies a positive thermal expansion (PTE) of the out-of-plane lattice constant \textit{c} and a weak negative thermal expansion (NTE) in the other crystallographic directions (Fig. \ref{fig: Low-temperature XRD}(b)).

The temperature dependence of the in-plane and out-of-plane lattice parameters shows opposite trends with respect to each other (Fig. \ref{fig: Low-temperature XRD}(b)). The net thermal expansion of unit-cell volume, defined as $(V_{300 K} - V_{20 K})/V_{20~K}$,
is positive and equals 0.757(6)\%. The change in lattice parameters (\textit{a}, \textit{c}) with temperature and unit-cell volume follow the Gr\"{u}neisen rule defined at zero pressure \cite{PhysRevB.54.R756}. This opposite temperature-dependent behaviour of the different lattice parameters reflects the effect of a strong distortion of the crystal lattice by FM-AFM transition~\cite{PhysRevB.106.094407}. 
Note that an anomaly of lattice constant around 190~K (similar to the sharp enhancement of magnetization with lowering temperature  as shown in Fig. \ref{fig: XRD_magnetization}(f) at 197~K) indicates a possible first-order-like magnetic transition where the strength of SPC is very high, as discussed later. 
The opposite temperature dependence of the lattice constants \textit{a} (negative) and \textit{c} (positive) between 20~K and 300 K eventually leads to anisotropic thermal expansion along crystallographic directions expressed as  $(a_{300~K} - a_{20~K})/a_{20~K}=-0.236(0)\%$ and $(c_{300~K} - c_{20~K})/c_{20~K}=1.235(0)\%$. It is evident from this comparison that a large anisotropy in thermal expansion is present. Since \textit{tr}-Cr$_{1.22}$Te$_{2}$ is an intermetallic compound, the total thermal expansion comprises both magnetic and non-magnetic parts. The non-magnetic thermal expansion further contains electronic and phononic contributions. Therefore, the total thermal expansion of any lattice parameter ($\eta$) can be expressed as $\eta=\eta_{mag}+\eta_{ele}+\eta_{phon}$, where $\eta_{mag}$ denotes the magnetic contribution. $\eta_{ele}$ represents the electronic contribution directly related to the density of electronic states at the Fermi level \cite{doi:10.1098/rspa.1956.0187}. The phonon contribution ($\eta_{phon}$) is estimated using second-order Debye-Gr\"{u}nesien formula at zero pressure \cite{Ranjan2011MagneticSA}:

\begin{equation}
\label{Phonon}
\eta[T]=\eta_{0}+\eta_{0}\frac{{U}}{Q-BU}
\end{equation}
where $\eta_{0}$ is the corresponding lattice parameter at absolute zero temperature, and internal energy ($U$) is expressed using the Debye approximation:
\begin{align}
\label{eq: Debye-Einstein equation}
U(T) =9Nk_{B}T\left(\frac{T}{\theta_D} \right)^3\int_{0}^{\theta_D/T} \frac{x^3}{(e^x - 1)}\,dx
\end{align}
where $N$ is the number of atoms per formula unit, and $\theta_D$ is the Debye temperature.

\begin{figure*}
      \centering
      \includegraphics[width=0.97\textwidth]{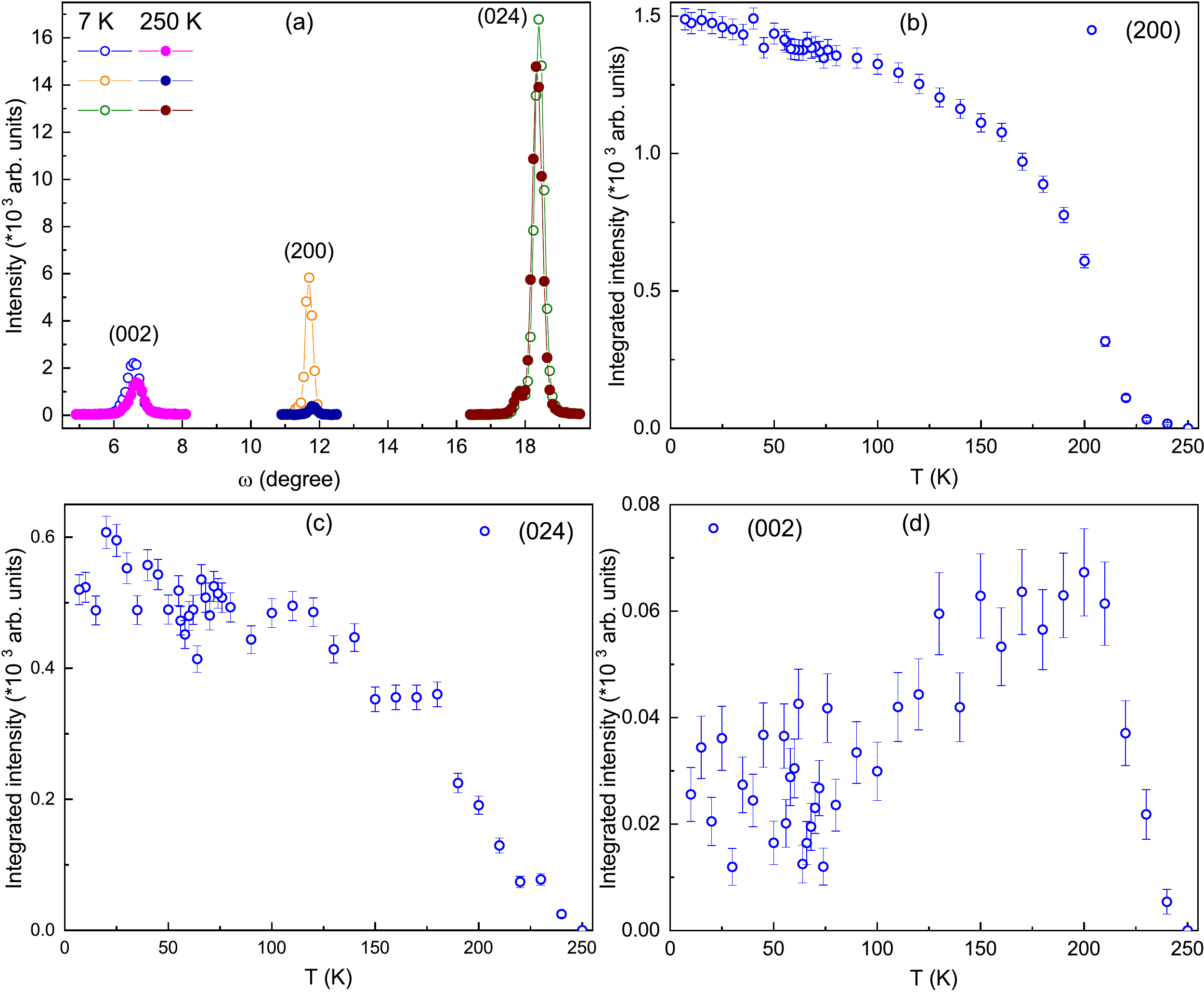}
      \caption{(a) The observed intensity vs temperature of the major commensurate peaks at 7~K and 250~K. (b), (c) and (d) are the temperature evolution of the integrated intensity of major commensurate peaks (200), (024), and (002) reflections, respectively, with respect to the same measured at 250~K. Here, (200) and (024) peaks show behaviour obtained from $M$ vs $T$ for $H\parallel$ \textit{c}-directions, whereas (002) indicates the trend similar to the $M$ vs $T$ for $H\perp$ \textit{c}-direction.}
      \label{fig: Magnetic peaks}
\end{figure*}



The effect of temperature results in thermal expansion behavior due to the asymmetric potential distribution of atoms, which results in positive thermal expansion of the lattice constant \textit{c}. However, atypical potential energy distribution could facilitate a phonon mode in crystallographic \textit{a(=b)}-axis, leading to nearly zero thermal expansion (NZTE) behaviour. This anomalous thermal behaviour has been realised in similar compounds with varying Cr and Te concentrations \cite{Li2022DiverseTE}. The presence of NZTE and the magnetic short-range correlations indicate the presence of SPC~\cite{10.1063/1.4914134}. Although SPC is an intrinsic property, its effect usually can be seen near magnetic transition, e.g., magnetostriction (change in the lattice caused by spin order).

The unit cell volume (\textit{V}) decreases with the lowering of temperature till 100~K, and below that, an NZTE is observed as shown by the open red circles in Fig. \ref{fig: Low-temperature XRD}(c). Furthermore, the Debye-Gr\"{u}neisen parameter has been fitted using Eq. \ref{Phonon} to model the phonon contribution, as shown by the solid blue line in the same Fig. \ref{fig: Low-temperature XRD}(c). The \textit{c}-axis follows the same trend as the volume thermal expansion and also shows an anomaly at 190 K (Fig. \ref{fig: Low-temperature XRD}(a)).
 Notably, both \textit{a} and \textit{b}-axes are still showing shortening above the \textit{T}$_\mathrm{C}$, which may be due to disordered Cr vacancies in the basal plane that counteract the contribution of the increased atomic vibrational amplitude to thermal expansion on heating. 
Based on the above discussion, it can be inferred that the thermal expansion behaviour is closely linked to magnetism and a particular Raman active mode (P2) is responsible for the observation of NZTE in \textit{tr}-Cr$_{1.22}$Te$_{2}$ as discussed later.
Many magnetic materials have been found to exhibit NZTE behaviours originating from the SPC, such as the Invar alloys of Fe-Ni, (Zr,Nb)Fe$_2$, Tb(Co,Fe)$_2$ \cite{Mohn1999}.
 
To reveal the relationship between the NZTE and magneto-volume effect (MVE), where normal positive thermal expansion is compensated by spontaneous magnetization~\cite{PhysRevB.59.269, PhysRevLett.75.3541}, in \textit{tr}-Cr$_{1.22}$Te$_{2}$, we calculated the spontaneous volume magnetostriction ($\omega_{s}$) with the following formula, 
\begin{equation}
\label{TE}
\omega_{s}=\omega_{exp}-\omega_{nm}
\end{equation}
Here, $\omega_{s}$ (blue solid squares in Fig. \ref{fig: Low-temperature XRD}(d)) is a parameter to physically quantify the contribution of MVE to the anomalous change in thermal expansion. $\omega_{exp}$ (black solid squares in Fig. \ref{fig: Low-temperature XRD}(d)) is the experimental data obtained from temperature-dependent PXRD, and $\omega_{nm}$ (red dashed line in Fig. \ref{fig: Low-temperature XRD}(b)) represents the normal thermal expansion derived from high-temperature behaviour. Meanwhile, according to the MVE theory \cite{RevModPhys.66.25, WOHLFARTH1980797, PhysRevB.29.5279}, $\omega_{s}$ has the following relationship with the local magnetic moments:
\begin{equation}
\label{ATE}
\omega_{s}=kCM(T)^2
\end{equation}
where \textit{k} is the compressibility, \textit{C} is the magnetovolume coupling constant, and \textit{M(T)} represents the local magnetic moments as a function of temperature. Fig. \ref{fig: Low-temperature XRD}(e) shows $\omega_{s}$ changes linearly with magnetization, which is replaced by local magnetic moment in a first approximation. Thus, MVE induces anomalous thermal expansion $\omega_s$, which minimises normal thermal expansion and produces NZTE at low temperatures. 

\subsubsection{Temperature-dependent Raman spectroscopy study}
The Raman active longitudinal optical phonon modes, $\Gamma_{Raman}$ =  9A$_{1\mathrm{g}}$+12E$_\mathrm{g}$ are allowed by the centrosymmetric point group D$_\mathrm{3d}$($\bar{3}$m) for the space group P$\bar{3}$m1 possessed by  \textit{tr}-Cr$_{1.22}$Te$_{2}$. A total of four peaks were observed in the unpolarized Raman data at RT (Fig. \ref{fig: LTRaman}(a)). The maximum intensity peaks (P2 and P3) appear at 120 cm$^{-1}$ and 138.4 $^{-1}$ for E$_\mathrm{g}$, and A$_{1\mathrm{g}}$ modes, corresponding to the in-plane vibrations and out-of-plane stretching, respectively. This is consistent with the reported \cite{Meng2021} Raman data where the peaks appear at 123.4 cm$^{-1}$ and 143.6 cm$^{-1}$. Shifting to the lower wavenumber side in our \textit{tr}-Cr$_{1.22}$Te$_{2}$ sample is possibly due to the modification of the phonon frequencies by excess Cr intercalation. The contour plot presented in Fig. \ref{fig: LTRaman}(b) shows the increased intensity of the P2 and P3 peaks around \textit{T}$_\mathrm{C}$, suggesting that the E$_\mathrm{g}$ and A$_{1\mathrm{g}}$ modes are strongly coupled with the magnetism (Fig. \ref{fig: LTRaman}(b)). Since both the P2 and P3 peaks are asymmetric, a Fano lineshape (Eq. \ref{Fano}) was implemented to deconvolute them \cite{PhysRev.124.1866}:
\begin{equation}
\label{Fano}
I(\omega) = I_0 \frac{(q + \epsilon)^2}{1 + \epsilon^2}
\end{equation}
where
$\epsilon = \frac{\omega - \omega_0}{\Gamma}$ with $\omega_{0}$ is the Fano resonance frequency, $q$ is the asymmetry  and $\Gamma$ represents Fano linewidth. It is important to note that the asymmetry parameter $q$ is directly coupled with the strength of the SPC constant ($\lambda_{SLC}$) as $1/q \propto \lambda_{SLC}$ \cite{RevModPhys.82.2257}. Raman shift and linewidth (FWHM) follow similar trends as a function of temperature for both maximum intensity peaks, corresponding to E$_\mathrm{g}$- and A$_{1\mathrm{g}}$-modes (Fig. \ref{fig: LTRaman}(c-d)). With decreasing temperature, a blueshift in both Raman mode frequencies indicates phonon hardening. This indicates that the frequency of the phonon, interacting with the incident photon, increases with lowering temperature down to 88~K with a clear anomaly around \textit{T}$_\mathrm{C}$. Just below 200~K, phonons soften, and peak-like features appear with the lowering of temperature. This is often seen in ferromagnetic systems and confirms the dominant FM interactions \cite{10.1063/1.2721142}. On the contrary, FWHM decreases with temperature reduction and broadens as the \textit{T}$_\mathrm{C}$ is approached. This is due to the strong scattering effect with increasing spin fluctuations as the magnetic correlation diverges. The temperature dependence of softening and FWHM broadening near \textit{T}$_\mathrm{C}$ (critical region) matches a typical scenario where phonons couple to magnetic order. Interestingly, the $1/q$ vs $T$ plot for both modes shows a peak-like feature around \textit{T}$_\mathrm{C}$ (Fig. \ref{fig: LTRaman}(e)). Here, an increase in the line shape asymmetry decreases the $q$ value, resulting in an increase of $1/q$. This can be inferred as the Fano resonance of discrete E$_\mathrm{g}$- and A$_{1\mathrm{g}}$-modes with the magnetic continuum strongly coupled to SPC in the system around \textit{T}$_\mathrm{C}$. An additional low-temperature anomaly in phonon softening and FWHM broadening near and below 100~K was observed for the E$_\mathrm{g}$-mode only. From the corresponding $1/q$ vs $T$ plot, the E$_\mathrm{g}$-mode shows a peak around 100~K, indicating the presence of strong SPC. It is important to note that the observation of NZTE from temperature-dependent PXRD patterns below 100~K is closely related to the in-plane vibrations of Cr-Te atoms, i.e., E$_\mathrm{g}$-mode. Furthermore, this 100~K anomaly correlating with the magnetization may indicate that further magnetic reorientation transition causes magnon excitations to modify the Raman intensity. Thus, SPC may possibly cause the NZTE at low temperatures. However, a more precise study with polarized Raman measurement and lattice dynamic calculation is needed, which is beyond the scope of this paper. This finding manifests the presence of SPC, which is strongly coupled with the magnetism in this system.

At high-temperature limit, i.e., in the paramagnetic state, lattice anharmonicity increases by the thermal energy \cite{PhysRevB.28.1928, PhysRevB.29.2051}. Therefore, to estimate the anharmonic dependence of the phonon frequency due to phonon-phonon interaction, Eq. \ref{Anharmonic} has been adapted: 
\begin{equation}
\label{Anharmonic}
\omega_{anh} (T) = \omega_0 - C\left[1+\frac{2}{(e^\frac{\hbar\omega}{k_BT}-1)}\right]
\end{equation}
Here, $\omega_0$ is the peak position extrapolated to zero temperature, and \textit{C} is a constant. The phonon renormalization is directly proportional to the spin-spin correlation function, $\langle \mathbf{S}_i \cdot \mathbf{S}_{i+1} \rangle$, between two nearest-neighbour spins located at $i$th and $(i+1)$th sites and expressed as $\Delta\omega$ = $\omega_{anh} (T) - \omega_0$ = $\lambda_{SLC}\langle \mathbf{S}_i \cdot \mathbf{S}_{i+1} \rangle \propto \frac{M^2(T)}{M_{max}^2}$ \cite{PhysRevB.60.11879}. Here, $\Delta\omega$ has been estimated from the Raman shift by fitting Eq. \ref{Anharmonic}. This is consistent with the magnetization measurement along the $c$-axis (Fig. \ref{fig: LTRaman}(f)). Thus, the abrupt change in magnetization at the \textit{T}$_\mathrm{C}$ is again inferred as the presence of strong SPC. Thus, low-temperature Raman measurements rule out the possibility of a structural phase transition by confirming lattice stability and the presence of strong SPC in this compound.

 \begin{figure*}
      \centering
      \includegraphics[width=0.97\textwidth]{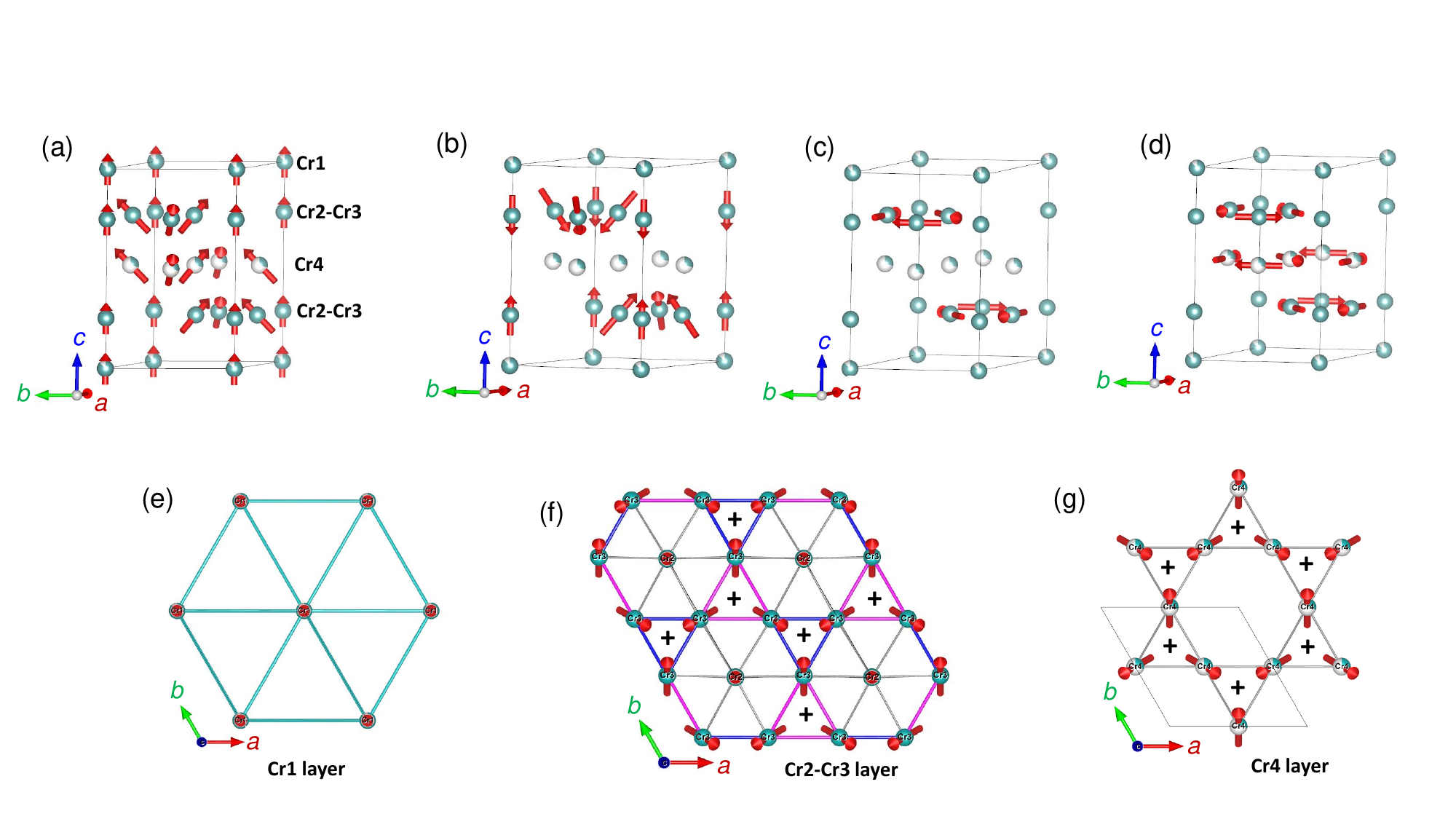}
      \caption{(a-d) (top panel) Magnetic spin configurations corresponding to the four magnetic subgroups P$\bar{3}$m$^\prime$1 (164.89), P$\bar{3}^{\prime}$m$^{\prime}$1 (164.88), P$\bar{3}$m$^{\prime}$1 (164.87), and P$\bar{3}$m1 (No. 164.85) for \textit{tr}-Cr$_{1.22}$Te$_{2}$. (e-g) (bottom panel) The spin configurations in the \textit{ab}-plane for different Cr-layers corresponding to the spin structure of (a). The value $\kappa$=+1(positive vector chirality) for all the spins points directly towards or away from the center of the triangle. Note that to illustrate the in-plane moment components and spin configurations in a better way, we have considered a larger value of M$_x$ for the schematic representations shown here.}
      \label{fig: Spin structures}
\end{figure*}

\begin{table}
    \centering
    \begin{tabular}{c| c| c| c| c}
        \toprule
        \textbf{Site} & \multicolumn{4}{c}{\textbf{Magnetic moment}} \\
        \cmidrule{2-5}
        & \textbf{P$\bar{3}$m$^\prime$1} & \textbf{P$\bar{3'}$m$^\prime$1} & \textbf{P$\bar{3'}$m1} & \textbf{P$\bar{3}$m1}\\
        & \textbf {(164.89)} & \textbf{(164.88)} & \textbf{(164.87)} & \textbf{(164.85)} \\
        \tabularnewline
        \hline
        Cr1 & (0,0,M$_z$) & (0,0,0) & (0,0,0) & (0,0,0) \\
        Cr2 & (0,0,M$_z$) & (0,0,M$_z$) & (0,0,0) & (0,0,0) \\
        Cr3 & (M$_x$,-M$_x$,M$_z$) & (M$_x$,-M$_x$,M$_z$) & (M$_x$,-M$_x$,0) & (M$_x$,M$_x$,0) \\
        Cr4 & (M$_x$,2M$_x$,M$_z$) & (0,0,0) & (0,0,0) & (M$_x$,0,0) \\
        \bottomrule
    \end{tabular}
    \caption{Magnetic moments at different sites for various space groups~\cite{AroyoPerezMatoCapillasKroumovaIvantchevMadariagaKirovWondratschek+2006+15+27, Aroyo2006a, Aroyo2006b}.}
    \label{tab:magnetic_moments}
\end{table}

\subsection{Magnetic ground state: Single crystal neutron diffraction study}
To determine the magnetic ground state, we performed the single crystal neutron diffraction experiment on a \textit{tr}-Cr$_{1.22}$Te$_{2}$ single crystal over the temperature range 7-290~K in zero field (inset of Fig. \ref{fig: Neutron refinement}(a)). A total of 138 reflections were measured at 290~K and 7~K (paramagnetic state) using a $^3$He point detector to validate the crystal and magnetic structures (Figs. \ref{fig: Neutron refinement}(a) and (b)). Out of these, a total 128 numbers of symmetry-independent (unique) reflections are considered for analysis at 290~K (Fig. \ref{fig: Neutron refinement}(a)). The single crystal diffraction data were analyzed starting with the crystal structure with the P$\bar{3}$m1 space group that was obtained from powder x-ray diffraction (Fig. \ref{fig: XRD_magnetization}(a)). The fitted integrated intensities and the agreement factors are shown in Fig. \ref{fig: Neutron refinement}(a). The refinement indicates that the Cr2 (6\textit{i}; (\textit{x}, -\textit{x}, \textit{z})) and Cr3 (2\textit{c};( 0, 0, \textit{z})) sites are fully occupied within the Cr-layer. The other two sites, Cr1 (1\textit{a}; (0, 0, 0)) and Cr4 (3\textit{f}; (1/2, 0, 1/2), are partially occupied with the occupancy of 0.891(6) and 0.293(4) respectively. It is important to remember that the occupancy of the Cr4 (interlayer) site can vary a lot due to the self-intercalation process, highly depending upon the growth conditions. The refined atomic positions of Cr and Te are similar to those obtained from PXRD at RT.

To find magnetic reflections, the sample was cooled to 7~K and a large region of reciprocal space was scanned with a 2D detector. Only the commensurate peaks were found at 7~K, suggesting the magnetic propagation vector k = (0 0 0). The intensity of the same 138 reflections was measured using the point detector at 7~K, and out of this, a total of 120 numbers of symmetry-independent (unique) reflections are considered for analysis (Fig. \ref{fig: Neutron refinement}(b)). The magnetic intensities were then determined from the difference in intensities between 7~K and 290~K. A few selected Bragg peaks measured at 7~K and 250~K are shown in Fig. \ref{fig: Magnetic peaks}(a). It may be noted that the peaks with indices (\textit{h}00), i.e., (200), show a strong enhancement of the intensity, followed by the peaks with mixed indices (0\textit{k}\textit{l}), i.e., (024). The enhancement of intensity for the (00\textit{l}) peaks, i.e., (002), is found to be relatively weak. Since neutron diffraction measures the moment component perpendicular to the momentum transfer, the observed intensity distributions described above reveal a major moment component along the \textit{c}-axis. 

To determine the nature of the magnetic state, the temperature dependence of a few selected magnetic Bragg peaks was investigated over the 7-290~K range with finer temperature steps. The temperature dependences of the integrated intensities of strong magnetic reflections, corresponding to the Miller indices (200), (024), and (002), are shown in Fig. \ref{fig: Magnetic peaks}(b-d). It may be noted that the temperature variation of the (200) magnetic Bragg peak intensity increases monotonously and resembles the susceptibility curve [$\chi=M/\mu_0H$ vs \textit{T}] for the applied magnetic field $H \parallel$\textit{c}  (Fig. \ref{fig: XRD_magnetization}(f)). On the other hand, the temperature dependence of the integrated intensity of the (002) magnetic Bragg peak follows the susceptibility curve [$\chi=M/\mu_0H$ vs \textit{T}] measured for the applied field $H \perp$ \textit{c}, where the intensity increases initially and reaches a maximum at 200~K and then decreases with further lowering of temperature down to 7~K. However, the observed intensity change at the (002) reflection could arise from a structural effect, such as magnetoelastic coupling as discussed in the previous sections, which might lower the local magnetic symmetry and induce weak structural distortions. The above observations suggest that ferromagnetic correlations develop for $H \parallel$\textit{c}-axis, whereas antiferromagnetic correlations appear in $H\perp$ \textit{c}-direction. It should be mentioned that the survival of finite intensity associated with the (200), (002) and (024) reflections even above \textit{T}$_\mathrm{C}$ indicates the presence of short-range correlations, which further supports the presence of geometrical frustration in the present compound.  

To determine the magnetic structure, we performed the magnetic space group (MSG) analysis using the MAXMGN program from the Bilbao Crystallographic server~\cite{AroyoPerezMatoCapillasKroumovaIvantchevMadariagaKirovWondratschek+2006+15+27, Aroyo2006a, Aroyo2006b}. The parent crystal structure with space group P$\bar{3}$m1 (space group No. 164) has been considered with all four Cr magnetic sites [Table.\ref{tab: magnetic structure data}]. The possible magnetic structures for the magnetic propagation vector k = (0 0 0) correspond to four MSG P$\bar{3}$m$^\prime$1 (magnetic space group No. 164.89), P$\bar{3'}$m$^\prime$1 (No. 164.88), P$\bar{3'}$m1 (No. 164.87), and P$\bar{3}$m1 (No. 164.85), which allow non-zero magnetic moments for at least one atom.

\begin{table}[h]
\begin{center}
\begin{adjustbox}{width=\columnwidth}
\begin{tabular}{|c|c|}
\hline
\textbf{Property} & \textbf{Details} \\
\hline
Parent Space Group & P$\bar{3}$m1 (No. 164) \\
\hline
Propagation Vector(s) & (0, 0, 0) \\
\hline
Transformation from Parent Basis & $(a, b, c; 0, 0, 0)$ \\
\hline
MSG Symbol & P$\bar{3}$m$^\prime$1 \\
\hline
MSG Number & 164.89 \\
\hline
Transformation to Standard MSG Setting & $(a,b,c;0,0,0)$ \\
\hline
Magnetic Point Group & m$^\prime$m$^\prime$m (2a+2b, c, -b) \\
\hline
Unit-cell Parameters & 
$a = 7.81530$ \AA, $\alpha = 90^\circ$ \\
& $b = 7.81530$ \AA, $\beta = 90^\circ$ \\
& $c = 11.9807$ \AA, $\gamma = 120^\circ$ \\
\hline
\multicolumn{2}{|c|}{\textbf{MSG Symmetry Operations}} \\
\hline
$x,y,z,+1$                & $\left\{1\;|\;0\right\}$ \\
$-y,x-y,z,+1$           & $\left\{3^+_{001}\;|\;0\right\}$ \\
$-x+y,-x,z,+1$        & $\left\{3^-_{001}\;|\;0\right\}$ \\
$-x,-y,-z,+1$           & $\left\{-1\;|\;0\right\}$ \\
$y,-x+y,-z,+1$        & $\left\{-3^+_{001}\;|\;0\right\}$ \\
$x-y,x,-z,+1$           & $\left\{-3^-_{001}\;|\;0\right\}$ \\
$x-y,-y,-z,-1$          & $\left\{2^\prime_{100}\;|\;0\right\}$ \\
$y,x,-z,-1$               & $\left\{2^\prime_{110}\;|\;0\right\}$ \\
$-x,-x+y,-z,-1$       & $\left\{2^\prime_{010}\;|\;0\right\}$ \\
$-x+y,y,z,-1$          & $\left\{m^\prime_{100}\;|\;0\right\}$ \\
$-y,-x,z,-1$             & $\left\{m^\prime_{110}\;|\;0\right\}$ \\
$x,x-y,z,-1$             & $\left\{m^\prime_{010}\;|\;0\right\}$ \\
\hline
\multicolumn{2}{|c|}{\textbf{Positions of Magnetic Atoms}} \\
\hline
Cr1 & $x=0.00,\;y=0.00,\;z=0.00$ \\
\hline
Cr2 & $x=0.00,\;y=0.00,\;z=0.2560$ \\
\hline
Cr3 & $x=0.5043,\;y=0.4957,\;z=0.2476$ \\
\hline
Cr4 & $x=0.50,\;y=0.00,\;z=0.50$ \\
\hline
\multicolumn{2}{|c|}{\textbf{Positions of Non-Magnetic Atoms}} \\
\hline
Te1 & $x=0.33333,\;y=0.66667,\;z=0.8834$ \\
\hline
Te2 & $x=0.33333,\;y=0.66667,\;z=0.3690$ \\
\hline
Te3 & $x=0.8328,\;y=0.1672,\;z=0.3779$ \\
\hline
Te4 & $x=0.1672,\;y=0.8328,\;z=0.1249$ \\
\hline
\multicolumn{2}{|c|}{\textbf{Magnetic Moments ($\mu_B$)}} \\
\hline
Cr1: $(0,\;0,\;m_z)$ & $(0,\;0,\;2.108)$ \\
& Magnitude = 2.108 \\
\hline
Cr2: $(0,\;0,\;m_z)$ & $(0,\;0,\;2.108)$ \\
& Magnitude = 2.108 \\
\hline
Cr3: $(m_x,\;-m_x,\;m_z)$ & $(-0.08,\;0.08,\;2.108)$ \\
& Magnitude = 2.111 \\
\hline
Cr4: $(m_x,\;2m_x,\;m_z)$ & $(-0.08,\;-0.16,\;2.108)$ \\
& Magnitude = 2.116 \\
\hline
\end{tabular}
\end{adjustbox}
\caption{List of magnetic symmetry operations and magnetic structural data.}
\label{tab: magnetic structure data}
\end{center}
\end{table}

The magnetic space group P$\bar{3}$m$^\prime$1 allows the finite magnetic moments on all four Cr sites [Table~\ref{tab:magnetic_moments} and Fig. \ref{fig: Spin structures}(a)]. Whereas the other subgroups allow finite magnetic moments either only on two Cr sites (P$\bar{3'}$m$^\prime$1 and P$\bar{3}$m1) or only on one Cr site (P$\bar{3}$m$^\prime$1) [Table- \ref{tab:magnetic_moments} and Fig. \ref{fig: Spin structures}(b-d)]. For the magnetic space group (P$\bar{3'}$m1; 164.87), the allowed magnetic moments are only for the Cr3 site, and they are confined within the \textit{ab}-plane. Such a magnetic structure does not correspond to the experimental observation of strong intensities for (\textit{h}\textit{k}0) magnetic Bragg peaks and weak intensities for magnetic Bragg peaks with indices (00\textit{l}); hence, it can be excluded.  For the magnetic space group (P$\bar{3}$\textit{m}1; 164.85) as well, the allowed magnetic moments on the Cr3 and Cr4 sites are confined within the \textit{ab}-plane, which do not match the experimental observations and, hence, can also be excluded.  The experimental pattern, measured at 7 K, was analysed with the magnetic structures corresponding to the subgroups P$\bar{3}$m$^\prime$1 (164.89) and P$\bar{3'}$m$^\prime$1 (164.88). A better fit was obtained for the magnetic structure corresponding to the magnetic space group P$\bar{3}$m$^\prime$1 (164.89). The fit between the observed and calculated $F^2$ at 7 K is shown in Fig. \ref{fig: Neutron refinement}(b). The refined magnetic moment components are M$_x$=0.08(8) $\mu_B$ and M$_z$=2.11(10) $\mu_{B}$. It is important to note that all four Cr sites are symmetry-inequivalent; the refinement should, in principle, allow for different moment magnitudes. However, for simplicity and better control of the refinement, the moments were constrained to be equal (i.e., equal M$_x$ value for Cr3 and Cr4 sites; and equal M$_z$ value for all four sites). Thus, our single crystal neutron diffraction experiment reveals the magnetic structure of \textit{tr}-Cr$_{1.22}$Te$_{2}$, characterised by a dominant moment component along the \textit{c}-axis and a small component in the \textit{ab}-plane. Here, we mention that the refined value of the M$_x$ components is small; a large error bar is associated with it. As other measurements, such as magnetization and THE, suggest a finite in-plane moment component. Further, the magnetic space group P$\bar{3}$m$^\prime$1 yields a possible M$_x$ component for Cr3 and Cr4 sites. We have, therefore, considered the M$_x$ component in our model and refined it to describe the ferromagnetic components. Thus, we acknowledge the experimental limitations in detecting the in-plane moment components. Here, we would like to mention that none of the maximal magnetic space groups permit the (002) Bragg reflection. Rather, the in-plane components within the P$\bar{3}$m$^\prime$1 model are considered from the magnetic scattering at the (110) reflection. The moments of the Cr atoms located at the edges of the unit cell point along the \textit{c}-axis, and the moments of the Cr atoms located at the triangular sites within the \textit{ab}-plane make an umbrella-like spin arrangement (Fig. \ref{fig: Spin structures}(a)). Such a canted ferromagnetic ground state is in agreement with some reported $ab-initio$ study~\cite{bigi2024bilayerorthogonalferromagnetismcrte2based, PhysRevMaterials.4.114001,PhysRevB.106.094407}. 

Fig. \ref{fig: Spin structures}(e), (f), and (g) represent the spin configurations for individual layers of Fig. \ref{fig: Spin structures}(a) for better clarity. Interestingly, an umbrella-like spin structure of the Cr3 and Cr4 moments possesses a finite scalar spin chirality (SSC) and is defined as 
$\chi = \mathbf{S}_1 \cdot (\mathbf{S}_2 \times \mathbf{S}_3)$ for three spins~\cite{doi:10.1126/science.1058161, Takagi2023, WANG2022100971}. Also, the all-in all-out spin configuration in each corner-shared triangle produced by Cr3 and Cr4 gives rise to a positive vector spin chirality (VSC) and is defined as $\vec{\kappa} = \left[\frac{2}{3\sqrt{3}}\right] (\vec{S}_1 \times \vec{S}_2 + \vec{S}_2 \times \vec{S}_3 + \vec{S}_3 \times \vec{S}_1)$ (represented by "$\kappa$=+1" here in Fig. \ref{fig: Spin structures}(f,g))~\cite{Grohol2005, Pradhan2023}. The real-space spin chirality induces an emergent Berry curvature, which is an intrinsic mechanism for the observed THE in \textit{tr}-Cr$_{1.22}$Te$_{2}$.

The observed spin chirality can arise due to the presence of frustration (as found in the present study) in a centrosymmetric crystal structure~\cite{Hirschberger2019, PhysRevLett.129.137202, https://doi.org/10.1002/adma.201900264, https://doi.org/10.1002/adma.201600889}. It should be mentioned that a kagome lattice (produced by the Cr moments for \textit{tr}-Cr$_{1.22}$Te$_{2}$ as shown in Fig. \ref{fig: XRD_magnetization} (c-e)) also inherently possesses a finite DMI due to the breaking of local inversion symmetry, thus even in a global centrosymmetric crystal structure, a finite DMI can be seen~\cite{PhysRevB.66.014422, PhysRevB.73.214446,PhysRevLett.101.026405}. Hence, the observed spin structure with chirality is responsible for the presence of the THE in \textit{tr}-Cr$_{1.22}$Te$_{2}$. 


\section{Conclusion} In this work, the ground state structural and magnetic properties of \textit{tr}-Cr$_{1.22}$Te$_{2}$ have been studied by low-temperature PXRD, magnetization, temperature-dependent Raman spectroscopy, and single crystal neutron diffraction. Low magnetic field measurements along the \textit{c}-axis indicate the presence of two successive magnetic transitions, FM-AFM (\textit{T}$_\mathrm{C}$=197~K), and AFM-PM (\textit{T}$_\mathrm{N}$=211~K) in this system. The low-temperature crystal structure properties reveal strong SPC without any structural phase transition. Whereas, the magnetization measurement with field perpendicular to the \textit{c}-axis shows AFM ordering below 197~K with significantly lower atomic moment values implying strong magnetocrystalline anisotropy. From the Curie-Weiss analysis, dominant FM interactions in the principal directions were found. Moreover, the \textit{Rhodes Wohlfarth} Ratio (RWR) indicates the presence of itinerant electrons in the system. Zero-field single crystal neutron diffraction was used to determine the magnetic arrangement of the Cr ions, which revealed a canted ferromagnetic ground state. The Cr1 and Cr2 sites are ferromagnetically aligned along the \textit{c}-axis. On the other hand, the Cr3 and Cr4 moments are tilted with 3\degree and 5\degree angles, respectively, from the \textit{c}-axis, having finite moment components in the \textit{ab}-plane. Thus, the moments from both the Cr3 and Cr4 sites make an umbrella-like structure. Finally, the overall spin arrangement of \textit{tr}-Cr$_{1.22}$Te$_{2}$ is  a canted ferromagnetic structure. This very novel finding sheds light on the complex nature of the magnetic ordering of this compound. Moreover, the umbrella-like spin structure possesses a finite spin chirality and is hence responsible for the observation of the topological Hall effect. The effect of frustration is also discussed in order to achieve the novel magnetic state at low temperatures. 

\section{Acknowledgement} S.M.H. and M.M. would like to acknowledge the UGC-DAE Collaborative Research Scheme (Ref. No. CRS/2021-22/01/393) for funding. The neutron single-crystal diffraction experiment was performed on
ZEBRA diffractometer at the Swiss spallation neutron source SINQ, Paul Scherrer Institute, Villigen, Switzerland. P.R.B and O.Z acknowledge Swiss National Science Foundation (SNSF) grant No. 200020-182536 for financial assistance. N.K. acknowledges the Science and Engineering Research Board (SERB), India, for financial support through Grant Sanction No. CRG/2021/002747 and Max Planck Society for funding under the Max Planck-India partner group project. This research project utilized the instrumentation facilities of the Technical Research Centre (TRC) at the S. N. Bose National Centre for Basic Sciences under the Department of Science and Technology (DST), Government of India.


\bibliography{References}
\end{document}